\documentclass[twocolumn,prl,superscriptaddress]{revtex4-2} 
\usepackage[T1]{fontenc}

\usepackage{amsmath,amsfonts,bm,amssymb}
\usepackage{graphicx}  %epsfig,
\usepackage{color}
\usepackage{extarrows}
\usepackage{enumitem}
\usepackage{empheq}
\usepackage{tensor}

\usepackage{amsthm}

\theoremstyle{remark}

\newcommand{\ud}{{\mathrm d}}

\newcommand{\w}{\omega}

\newcommand{\B}{\mbox{\tiny B}}

\newcommand{\tS}{\mbox{\tiny S}}

\newcommand{\T}{\mbox{\tiny T}}

\newcommand{\SB}{\mbox{\tiny SB}}

\newcommand{\la}{\langle}
\newcommand{\ra}{\rangle}

\newcommand{\nl}{\nonumber \\}
\newcommand{\be}{\begin{equation}}
\newcommand{\ee}{\end{equation}}
\newcommand{\bsube}{\begin{subequations}}
\newcommand{\esube}{\end{subequations}}
\newcommand{\Eq}[1]{Eq.\,(\ref{#1})}
\newcommand{\Eqs}[1]{Eqs.\,(\ref{#1})}
\newcommand{\Fig}[1]{Fig.\,\ref{#1}}

\newcommand{\RN}[1]{%
  \textup{\uppercase\expandafter{\romannumeral#1}}%
}

\allowdisplaybreaks[1]

\definecolor{darkblue}{RGB}{0, 56, 102}

\usepackage{hyperref}  
\hypersetup{hidelinks,
	colorlinks=true,
	allcolors=black,
	pdfstartview=Fit,
	breaklinks=true
}

\begin{document}

\title{
The Phase-Coupled Caldeira-Leggett Model: Non-Markovian Open Quantum Dynamics beyond Linear Dissipation
}
%%%

\author{Ao-Xiang Chang}
\thanks{Authors of equal contributions}
\affiliation{
  Hefei National Research Center for Physical Sciences at the Microscale, University of Science and Technology of China, Hefei, Anhui 230026, China
}
\affiliation{
  Department of Modern Physics, University of Science and Technology of China, Hefei, Anhui 230026, China
}
\author{Yu Su}
\thanks{Authors of equal contributions}
\email{suyupilemao@mail.ustc.edu.cn}

\author{Zi-Fan Zhu}
\author{Yao Wang}
\email{wy2010@ustc.edu.cn}
\author{Rui-Xue Xu}
\author{YiJing Yan}
\email{yanyj@ustc.edu.cn}
\affiliation{
  Hefei National Research Center for Physical Sciences at the Microscale, University of Science and Technology of China, Hefei, Anhui 230026, China
}

\date{\today}

\begin{abstract}
We introduce the \textit{Phase-Coupled Caldeira-Leggett} (PCL) model of quantum dissipation and develop an exact framework for its dynamics. 
Unlike the conventional Caldeira-Leggett model with linear system-bath coupling $H_{\mathrm{SB}}\!\propto\!\hat F$, the PCL model features an exponential interaction $H_{\mathrm{SB}}\!\propto\! e^{i\lambda \hat F}$, where $\hat F$ denotes the collective bath coordinate. 
This model unifies concepts from quantum Brownian motion and polaron physics, providing a general platform to study phase-mediated dissipation and decoherence beyond the linear-response regime. 
Despite its nonlinear system-bath coupling, the Gaussian nature of the environment allows a nonperturbative and non-Markovian treatment of PCL model within the algebra of dissipative quasiparticles. 
We obtain an exact closed-form equation of motion for the reduced density operator, and numerical simulations reveal distinctive dynamical behaviors that deviate markedly from those predicted by the conventional Caldeira--Leggett model.
\end{abstract}

\maketitle

\paragraph{Introduction.} Understanding dissipation and decoherence is a central challenge in quantum science, impacting fields ranging from condensed matter physics and quantum optics to quantum information processing. Open quantum systems, which interact with their environment, exhibit rich phenomena such as relaxation, noise-induced transitions, and decoherence, which fundamentally plays roles in developing techniques about the quantum coherence and control \cite{Wei21,Bre02,Scu97,Van051037,Ram98,She84, Bre16021002,Dev17015001,Wei21015008,Wan21142501,Sch23160402,Kum24070404,Lan24020201}. The Caldeira-Leggett (CL) model has long provided a paradigmatic framework for describing such effects, modeling a system linearly coupled to a bath of harmonic oscillators. This model successfully captures quantum Brownian motion, friction, and classical-to-quantum crossover behavior, and has served as a cornerstone for quantum dissipation theory \cite{Cal81211,Cal83587, Cal851059,Leg871}.

Despite its success, the standard CL model is restricted to \emph{linear} system–bath couplings, i.e.,
\be 
H_{\tS\B}=\hat S \hat F,
\ee
where  $\hat F = \sum_jc_j\hat x_j$ denotes the collective bath coordinate and $\hat S$ is a system operator. This type of coupling primarily captures the case that the environment linearly responses towards the reaction of system. 
A distinct and physically rich class of models emerge when the system-bath interaction is mediated via the \emph{exponential} of the environmental operator: 
\begin{align}\label{HSB} 
  H_{\SB} = \hat S \hat B, \quad\text{with}\quad \hat B = e^{i\lambda\hat F} + e^{-i\lambda\hat F}.
\end{align} 
Here, $\hat B$ is a bath operator generated by exponential of $\hat F$, with $\lambda$ a real parameter. 
This is referred to as the Phase-Coupled Caldeira-Leggett (PCL) model, in which the system couples to the environment through an exponential operator, $\hat B$, in a phase-dependent or polaron-like fashion. 
Such nonlinear couplings arise naturally in a variety of physical contexts, especially in those concerning with strong correlations and collective excitations, e.g., periodic quantum dissipative system \cite{Sch831506,Fis856190, Gui86263}, polaron dynamics \cite{Lee53297, Xu16110308}, and transport in Luttinger liquid \cite{Kan921220, Hat14115103}. In these scenarios, the system modulates the collective phase or displacement of the bath modes. 
Accurately solving the PCL model's dynamics is essential for understanding quantum transport, coherence revival, and decoherence mechanisms in these scenarios.

Over the past decades, a variety of theoretical frameworks have been developed to describe quantum dissipation, ranging from perturbative master equations under the Markovian approximation \cite{Blo571206, Red651, Leg871} to fully nonperturbative and non-Markovian approaches based on path integrals \cite{Fey63118, Mey9073, Mak954600, Mak954611, Str183322, Mak211, Mak20041104}. Among the latter, the Feynman-Vernon influence functional provides a powerful formalism for characterizing environmental effects through bath correlation functions \cite{Fey63118, Leg871}. 
For the conventional CL model, this functional admits an analytical expression governed by a memory kernel, forming the foundation of modern non-Markovian methods such as the hierarchical equations of motion (HEOM) \cite{Tan89101,Tan906676,Tan06082001,Yan04216,Xu05041103, Xu07031107,Tan20020901}. However, extending these techniques to PCL model poses major challenges: Directly applying Wick’s theorem leads to formidable algebraic complexity, preventing an exact evaluation of the influence functional \cite{Bog09, Kub621100, Fox78179,Mak992823,Zha25115131}. Consequently, existing studies of PCL model remain largely limited to perturbative and/or Markovian approximations.

\begin{figure*}[ht]
    % \centering
    \includegraphics[width=0.9\linewidth]{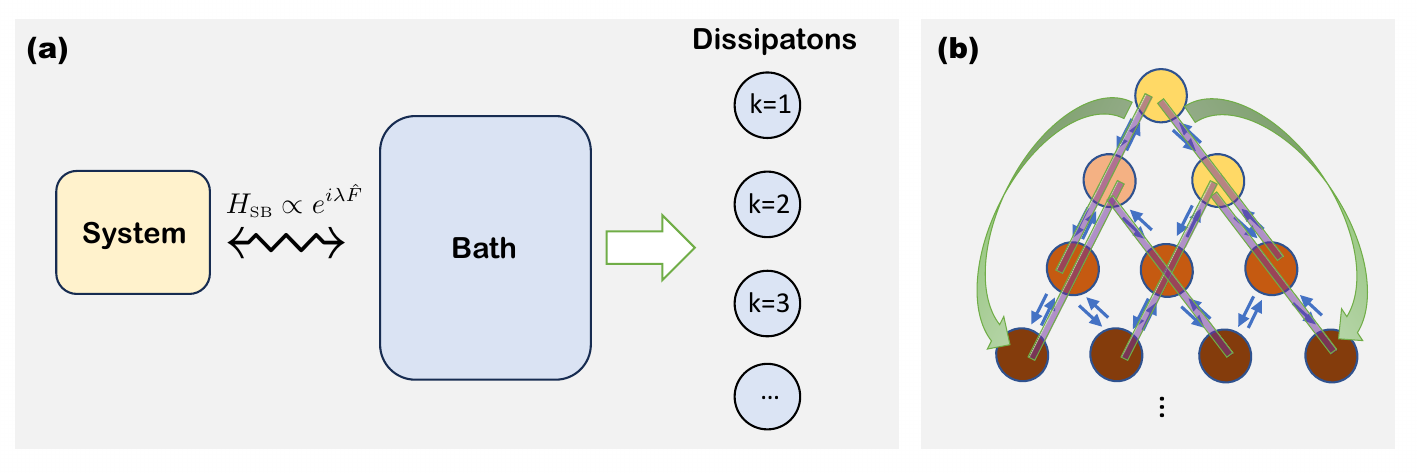}
    \caption{(a) Schematic illustration of the PCL model [cf.\,\Eq{HSB}] and DD of bath [cf.\,\Eq{map1}]. A quantum system 
interacts with a harmonic bath 
through an exponential, phase-type coupling. This interaction mediates dissipation through phase modulation rather than linear displacement, distinguishing the PCL model from the conventional CL framework. The  bath is further decomposed into dissipatons via the DD, preserving complete non-Markvian informations. (b) Illustration of the equations of motion \Eq{eom} in a hierarchical structure, showing the coupings across different levels of the dynamical varibales, including coupling of the nearest neighbor layer, next-nearest neighbor layer, next-next-nearest neighbor layer, and so on.}
    \label{fig1}
\end{figure*}

In this work, we develop an in-principle exact framework to obtain accurate, nonperturbative, and non-Markovian dynamics of the PCL model. This approach builds upon the \emph{dissipaton formalism} and its underlying algebra of statistical quasi-particles \cite{Yan14054105, Wan22170901, Jan25035111}. 
Originating from the intrinsic algebraic structure of HEOM, the dissipaton formalism offers a powerful route to extend open quantum system theory to nonlinear system-bath couplings \cite{Xu18114103, Che23074102, Su23024113, Su25_arxiv_2503_00297}.
The central idea of dissipaton formalism is the introduction of dissipatons-statistical quasi-particles that generalize the conventional Hilbert-space quasi-particles to Liouville space by extending Wick’s theorem into the complex plane \cite{Su25_arxiv_2503_00297, Jan25035111}. The key innovation of this work is the formulation of a \textit{generalized normal ordering} for dissipaton operators, which enables the construction of a closed algebra in terms of Hermite polynomials. The resulting \emph{equations of motion} exhibit a hierarchical structure that, while reminiscent of the HEOM, differ fundamentally in both structure and applicability. This exact equation of motion constitutes the core result of this work and provides the foundation for our subsequent numerical analysis.

\paragraph{Dissipaton formalism.} 
The total  Hamiltonian has the generic form of
$
H_{\T}=H_{\tS}+H_{\tS\B}+H_{\B}
$. 
Here,  $H_{\tS}$ and $H_{\B}$ are the system and bath Hamiltonian, respectively.
Our central goal is to solve the dissipative dynamics of PCL model, with the system-bath coupling in \Eq{HSB}.
We always set $\hbar\equiv 1$ and $\beta \equiv 1/(k_{B}T)$, where $k_{B}$ is the Boltzmann constant and $T$ the environment temperature.
For the PCL model, the environmental influence on the system is completely characterized by the bath correlation function, $C(t) = \langle \hat F(t)\hat F(0) \rangle_{\B}$, where $\hat F(t) \equiv e^{iH_{\B}t}\hat F e^{-iH_{\B}t}$ is defined in the bath Heisenberg picture, and the bath average $\langle\,\cdot\,\rangle_{\B}$ is taken over the equilibrium thermal state, $\rho_{\B}^{\mathrm{eq}}$. This fact follows directly from the structure of the influence functional: each term in its Dyson's expansion can be expressed as a product of $C(t)$ through Wick's theorem \cite{Zha25115131,Tam18030402}. Nevertheless, the total contributions from all cumulants are generally intractable to evaluate analytically \cite{Bog09, Kub621100, Fox78179,Mak992823}. To address this challenge, we employ the dissipaton formalism, whose underlying quasi-particle algebra provides a natural and efficient means to overcome this difficulty.

The formalism starts with the dissipaton decomposition (DD) of the bath collective coordinate [cf. \Fig{fig1}(a)], such that the correlation function $C(t)$ remains invariant. Thus, the non-Markovian information is fully realized through the statistics of the dissipatons. The DD maps $\hat F$ into $\{\hat f_k\}$, dissipaton operators \cite{Yan14054105},
\be \label{map1}
    \hat F \stackrel{\text{DD}}{\longmapsto}\sum_{k=1}^{K}\hat f_k.
\ee
All dissipatons are assumed to be mutually independent, satisfying the correlation relation $\langle \hat f_k(t) \hat f_{k'}(0) \rangle_{\B} = \delta_{kk'} c_k(t)$. Consequently,
\begin{align}\label{condition}
C(t) = \sum_{k,k'=1}^{K} \langle \hat f_k(t)\hat f_{k'}(0)\rangle_{\B} = \sum_{k=1}^{K} c_k(t).
\end{align}
In this work, we adopt $c_k(t)=\eta_k e^{-\gamma_k t}$, with $\eta_k$ and $\gamma_k$ being complex. Intuitively, the real and imaginary parts of $\gamma_k$ characterize the damping rate and oscillation frequency of the $k$th dissipaton mode, respectively. The exponential decomposition of $C(t)$ can be achieved for arbitrary spectral density at any temperatures, by employing numerical fitting schemes, such as the time-domain Prony fitting, the numerical analytic continuation, and so on \cite{Che22221102, Xu22230601, Zha24035154, Zha25214111}. Most of the decomposition methods lead to the exponents are either real or complex conjugate paired. Therefore, the backward correlation $C^*(t)$ shares the same exponents, that is, $C^*(t) = \la \hat F(0)\hat F(t)\ra_{\B} = \sum_k \eta_{\bar k}^*e^{-\gamma_kt}$. Here, we denote the index $\bar k$ by $\gamma_{\bar k}\equiv\gamma_{k}^{*}$. 

For establishing the non-Markovian dissipative dynamics, we study the collective dynamics of the system and dissipatons. To this end, we introduce the dynamical variables defined as
\begin{align}\label{DDO}
  \rho_{\bf n}^{(n)} \equiv \rho_{n_1\cdots n_K}^{(n)} \equiv {\rm tr}_{\B}\big[ \mathcal O(\hat f_1^{n_1}\cdots\hat f_{K}^{n_K})\rho_{\T} \big].
\end{align}
Here, $\rho_{\T}$ is the total density operator obeying the dynamics generated by $H_{\T}$ and ${\rm tr}_{\B}$ means the partial trace over the bath degrees of freedom. 
The index $n_k$ is a non-negative integer, representing the number of $k$-th dissipaton. We also denote $n \equiv \sum_k n_k$ as the total number of dissipatons in the superscript. The reduced density operator $\rho_{\tS}$ is just $\rho^{(n)}_{\bf n}$ with $n_1=\cdots =n_K=0$, that is $\rho^{(0)}_{\bf 0} = \rho_{\tS}\equiv {\rm tr}_{\B}(\rho_{\T})$. In \Eq{DDO}, the key concept is the the generalized normal ordering for dissipaton operators, denoted by $\mathcal O(\cdots)$, which will be specified below.

\paragraph{Generalized normal ordering.}  The generalized ordering is introduced to implement the intrinsic relation between different dynamical variables as defined in \Eq{DDO}. For practical use, we first present the properties that will be employed in constructing the equations of motion:
\begin{enumerate}[fullwidth, label=(\roman*)]
  \item The average of operators in the ordering over the bath thermal state is zero, i.e.,
  \be \label{zerott}
  {\rm tr}_{\B}\big[ \mathcal O(\hat f_1^{n_1}\cdots\hat f_{K}^{n_K})\rho_{\B}^{\rm eq} \big] = 0.
  \ee
  \item The bare-bath evolution is govern by generalized diffusion equation, 
\begin{align}\label{diff_eq}
    {\rm tr}_{\B}[\mathcal O(\dot{\hat f}_k)\rho_{\T}] = -\gamma_k{\rm tr}_{\B}[\mathcal O(\hat f_k)\rho_{\T}],
  \end{align}
  with $\dot{\hat f}_k \equiv i[H_{\B},\hat f_k]$.
  \item The generalized Wick's theorem: 
  \bsube\label{gwick}
  \begin{align}\label{wick1}
      {\rm tr}_{\B}\bigg[ \mathcal O\bigg( \prod_{k'} \hat f_{k'}^{n_{k'}} \bigg) \hat f_k^{>} \rho_{\T} \bigg] &= \rho_{{\bf n}_k^{+}}^{(n+1)} + n_k\eta_{k}^{}\rho_{{\bf n}_k^{-}}^{(n-1)},\\
      \label{wick2}
      {\rm tr}_{\B}\bigg[ \mathcal O\bigg( \prod_{k'} \hat f_{k'}^{n_{k'}} \bigg) \hat f_k^{<} \rho_{\T} \bigg] &= \rho_{{\bf n}_k^{+}}^{(n+1)} + n_k\eta_{\bar k}^{*}\rho_{{\bf n}_k^{-}}^{(n-1)}.
  \end{align}
  \esube
  Here, we denote the left and right action superoperators as $\hat f^{>}_k(\cdot) \equiv \hat f^{}_k(\cdot)$ and  $\hat f^{<}_k(\cdot) \equiv (\cdot)\hat f^{}_k$, respectively. %As a simple example, we have ${\rm tr}_{\B}(\hat f_k^{\gtrless}\rho_{\T}) = {\rm tr}_{\B}[\mathcal O(\hat f_k)\rho_{\T}]$. 
\end{enumerate}

Although these rules suffice to construct the equations of motion for the PCL model, we now comment on the detailed meaning of the generalized normal ordering. Indeed, each dissipaton operator can be decomposed into two components, $\hat f_k = \hat f_k^+ + \hat f_k^-$, in the sense of the thermofield mapping \cite{Ume95, Su25_arxiv_2503_00297}. Within this representation, the thermal bath state is effectively mapped onto a vacuum state, satisfying $\hat f_k^- \rho_{\B}^{\mathrm{eq}} = \rho_{\B}^{\mathrm{eq}} \hat f_k^+ = 0$. Consequently, the generalized normal ordering is defined by placing all $\hat f_k^+$ operators to the left of $\hat f_k^-$, which gives rise to the \Eqs{zerott} and (\ref{diff_eq}) \cite{Yan14054105}. Different from the conventional normal ordering, the generalized Wick's theorem concerns the contractions of $\hat f_k^>$ and $\hat f_k^<$ into the ordering, as seen in \Eq{gwick}, which reflects the difference between forward correlation $C(t)$ and backward correlation $C(-t)$ \cite{Yan14054105,Su25_arxiv_2503_00297}. The present formalism remains valid for the environments with discretized modes. For example, when $H_{\B} = \omega_0 (\hat p^2 + \hat x^2)/2$, the bath correlation $C(t)$ consists of two modes with purely imaginary, conjugate exponents $\pm i\omega_0$. The corresponding dissipaton operators are simply the creation and annihilation operators, $(\hat x \pm i \hat p)/\sqrt{2}$. In this case, the associated Wick’s theorem reduces to the conventional form \cite{Su25_arxiv_2503_00297,Jan25035111}.

\paragraph*{Hermite polynomial technique.}
For constructing the equations of motion  for the PCL model, we have to evaluate the Wick's contraction of $e^{i\lambda\hat F}\rightarrow  e^{i\lambda\sum_k\hat f_k}$. 
For simplicity, we illustrate the technique with a single dissipaton type, $\hat f$. The extension to multiple types is straightforward, as each dissipaton space is independent with the others.

Firstly, we introduce the Hermite polynomials for dissipaton operators, defined as  
\be \label{herm}
H_n^{>}(\hat f) \equiv \frac{\ud^n}{\ud z^n}e^{z\hat f - \eta z^2/2}\bigg|_{z=0} \ \ \text{with}\ \ n=0,1,2,\cdots.
\ee
Then there exist the relations
\begin{align}
  \mathcal O(\hat f^{n}) = H_{n}^{>}( \hat f)\ \text{ and }\ \hat f^{n} = i^{-n}\mathcal O[H_{n}^{>}(i\hat f)].
\end{align}
Using the generating function of Hermite polynomials, we obtain  \cite{Fan112145}
\begin{align}\label{10}
  e^{i\lambda\hat f} = \mathcal O\bigg[\sum_{n=0}^\infty\frac{\lambda^n}{n!}H_{n}^{>}(i\hat f)\bigg] = \mathcal O (e^{i\lambda\hat f - \eta\lambda^2/2}).
\end{align}
Since $e^{z_1\hat f - \eta z_1^2/2}e^{z_2\hat f - \eta z_2^2/2} = e^{(z_1+z_2)\hat f - \eta(z_1^2+z_2^2)/2}$, a similar procedure gives rise to
\begin{align}\label{11}
  \mathcal O(\hat f^m)\mathcal O(\hat f^n) =\!\! \sum_{l=0}^{\min(m,n)}{m\choose l}{n\choose l}\eta^ll!\mathcal O(\hat f^{m+n-2l}).
\end{align}
As a result, the contraction of $e^{i\lambda\hat f}$ is evaluated using \Eq{11} followed with \Eq{10}, resulting in
\begin{align}
  \mathcal O(\hat f^n)e^{i\lambda\hat f^>}\!\! = e^{-\frac{\eta\lambda^2}{2}}\!\!\sum_{m=0}^\infty\!\!\!\sum_{l=0}^{\min(m,n)}\!\!\frac{(i\lambda)^m\eta^l}{(m-l)!}{n\choose l}\mathcal O(\hat f^{n+m-2l}).
\end{align}
For the left action $\hat f^<$, we define $H_n^{<}(\hat f)$, with $\eta$ replaced by $\eta^*$ in \Eq{herm}, and the contraction relations are derived with a similar procedure.
\begin{figure*}[ht]
    % \centering
    \includegraphics[width=\textwidth]{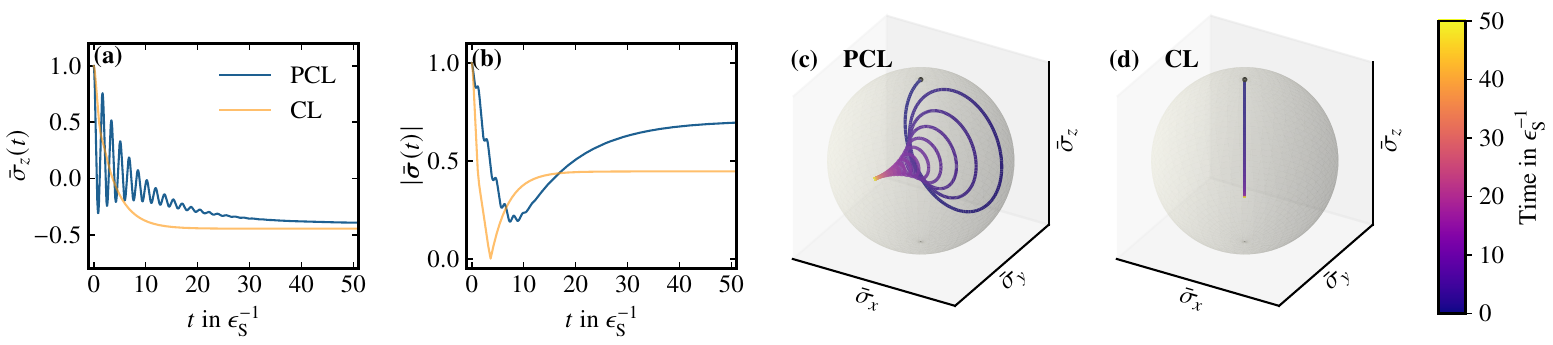}
    \caption{Numerical results for the two-level system dynamics under the PCL and CL system–bath interactions. The system Hamiltonian is $H_{\tS} = \epsilon_{\tS}\hat\sigma_z$ with the coupling operator $\hat S = \alpha\hat\sigma_x$. The bath is modeled by the Drude spectral density, $J(\omega) = \xi \omega / (\omega^2 + \gamma^2)$. The parameters are chosen as $\alpha = \epsilon_{\tS}$, $\gamma = \epsilon_{\tS}$, $k_BT = 2\epsilon_{\tS}$, $\xi = 1$, and $\lambda = 0.5$. Here, we evaluate the transient expectations of Pauli matrices, $\bar\sigma_i(t) \equiv {\rm tr}_{\tS}[\hat\sigma_i\rho_{\tS}(t)]$. For both the PCL and CL Hamiltonians, panels (a) and (b) show the population dynamics characterized by $\bar\sigma_z(t)$ and $|\bar{\bm\sigma}(t)| \equiv \sqrt{\bar\sigma_x^2(t) + \bar\sigma_y^2(t) + \bar\sigma_z^2(t)}$, respectively. Panels (c) and (d) depict the corresponding trajectories within the Bloch sphere for the PCL and CL cases. Remarkably, under the PCL interaction, the eigenvectors of the steady state $\lim_{t\to+\infty}\rho_{\tS}(t)$ deviate significantly from the eigenstates of $H_{\tS}$.}
    \label{fig2}
\end{figure*}

\paragraph*{Equations of motion.}
We are now ready to construct the equations of motion for PCL model. 
The strategy is straightforward. 
Starting from the Liouville--von Neumann equation in the total system--bath space,
\be 
\dot{\rho}_{\T}(t) = -i [H_{\tS} + H_{\B} + H_{\tS\B}, \rho_{\T}(t)],
\ee
we multiply both sides by $\mathcal{O}\!\left(\prod_k \hat f_k^{n_k}\right)$ 
and take the partial trace over the bath degrees of freedom. 
The system Hamiltonian term directly contributes $-i[H_{\mathrm{S}}, \rho_{\mathbf{n}}^{(n)}]$, 
while the bath Hamiltonian term is treated via the generalized diffusion equation [\Eq{diff_eq}], 
yielding the contribution $-\sum_k n_k \gamma_k \rho_{\mathbf{n}}^{(n)}$. 
The interaction term is evaluated using the generalized Wick’s theorem together with the properties of Hermite polynomials. 
After some straightforward but lengthy algebra, we arrive at the \emph{central result}---the 
equations of motion for the PCL model:
\begin{widetext}
    \begin{align} \label{eom}
   % &\dot\rho_{\bf n}^{(n)}= -\bigg( i\mathcal L_{\tS} + \sum_{k}n_k\gamma_k \bigg)\rho_{\bf n}^{(n)} 
  % -i g \prod_k\sum_{m_k=0}^\infty \sideset{}{'}{\sum}_{l_k=0}{n_k\choose l_k} 
      % \nl & 
  % \times \frac{(i\lambda)^{m}\!+\!(-i\lambda)^{m}}{(m_k-l_k)!}
  % \Big(\eta_k^{l_k}\hat S\rho_{\mathbf n+\mathbf m-2\mathbf l}^{(n+m-2l)} -\eta^{\ast l_k}_{\bar k}\rho_{\mathbf n+\mathbf m-2\mathbf l}^{(n+m-2l)}\hat S\Big).
  \dot\rho_{\bf n}^{(n)} &= -i[H_{\tS},\rho_{\bf n}^{(n)}] - \sum_k n_k \gamma_k \rho_{\bf n}^{(n)} - i g  \sideset{}{'}{\sum}_{\bf m,l} [(i\lambda)^m - (-i\lambda)^m] \prod_k \frac{\eta_k^{l_k}}{(m_k - l_k)!}{n_k\choose l_k}\hat S \rho_{\mathbf n+\mathbf m-2\mathbf l}^{(n+m-2l)} \nl
    &\quad\, + i g  \sideset{}{'}{\sum}_{\bf m,l} [(i\lambda)^m - (-i\lambda)^m] \prod_k \frac{\eta_{\bar k}^{*l_k}}{(m_k - l_k)!}{n_k\choose l_k}\rho_{\mathbf n+\mathbf m-2\mathbf l}^{(n+m-2l)}\hat S.
\end{align}
\end{widetext}
We present the derivations in Supplementary Material (SM) \footnote{See Supplemental Material for details for constructing the equations of motion.}.
Here, $g\equiv e^{-\lambda^2\la\hat F^2\ra_{\B}/2}$ is a real number with $\la\hat F^2\ra_{\B} \equiv {\rm tr}_{\B}(\hat F^2\rho_{\B}^{\rm eq})$. The prime summation is defined as ${\sum}'_{\bf m,l} \equiv \sum_{m_1=0}^\infty\cdots\sum_{m_K=0}^\infty \sum_{l_1=0}^{\min(m_1,n_1)}\cdots\sum_{l_K=0}^{\min(m_K,n_K)}$, which gives non-negative lower indices of $\rho_{\mathbf n+\mathbf m-2\mathbf l}^{(n+m-2l)}$ with $m = \sum_k m_k$ and $l = \sum_k l_k$. The initial conditions for \Eq{eom} are given by $\rho_{\bf 0}^{(0)}(0) = \rho_{\tS}(0)$ and $\rho_{\bf n}^{(n>0)}(0) = 0$ [cf.\,\Eq{zerott}]. The structure of \Eq{eom} in illustrated in \Fig{fig1}. Like propagating the HEOM, elaborating \Eq{eom} numerically needs a truncation of the label indices. One practical choice of tier-level truncation scheme, that is,  $\rho_{\bf n}^{(n>L)}=0$ with $L$ labeling the truncation level. The error of the propagation of the non-Markovian dynamics will decrease when we select a larger $L$; See SM for details.

\paragraph{Numerical illustration.} 
The interaction between the system and environment leads to the irreversible dynamical phenomena. Here, for discussing the physical influence of the PCL bath, we consider a simple two-level system, with $H_{\tS} = \epsilon_{\tS}\hat\sigma_z = \epsilon_{\tS}(|0\ra\la 0| - |1\ra\la 1|)$ and $\hat S = \alpha\hat\sigma_x = \alpha(|0\ra\la 1| + |1\ra\la 0|)$. Here, $\epsilon_{\tS}$ and $\alpha$ being the bare-system eigen-energy and system-bath coupling strength, respectively. Furthermore, we adopt the Drude model for the bath, $J(\omega) = \xi \omega / (\omega^2 + \gamma^2)$. The bath correlation  can be obtained via the fluctuation-dissipation theorem, 
\be 
C(t)=\frac{1}{\pi}\int_{-\infty}^{\infty}\!\!{\rm d}\w\, e^{-i\w t}\frac{J(\omega)}{1 - e^{-\beta\omega}}.
\ee
Within the numerical evaluation, we set $\xi = 1$, $\gamma=\epsilon_{\tS}$, and the temperature $k_BT = 2\epsilon_{\tS}$. Here, we select $K=2$ and $L=6$ to guarantee the accuracy of the propagation dynamics. 

% Figure \ref{fig2} illustrates the time evolution of the reduced density operator $\rho_{\tS}(t)$ for both the PCL and CL models. To visualize the system dynamics, we adopt the Bloch sphere representation, where each component is defined as $\bar\sigma_i(t) \equiv {\rm tr}{\tS}[\hat\sigma_i\rho_{\tS}(t)]$ for $i = x, y, z$. The trajectory of a mixed state is then represented as a path within a unit sphere. Panels (c) and (d) of \Fig{fig2} display the Bloch trajectories corresponding to the PCL and CL models, respectively, while panels (a) and (b) show the transient results of $\bar\sigma_z(t)$ and $|\bar{\bm\sigma}(t)| \equiv \sqrt{\bar\sigma_x^2(t) + \bar\sigma_y^2(t) + \bar\sigma_z^2(t)}$. The two models exhibit qualitatively distinct behaviors. (i) At short times, the PCL dynamics displays pronounced coherent oscillations with a high frequency, whereas the CL dynamics show a rapid monotonic decoherence. (ii) In the long-time limit, the steady system state of the CL model becomes diagonal in the eigenbasis of the system Hamiltonian, while that of the PCL model does not commute with the Hamiltonian; the corresponding eigenvectors $|\psi_{\pm}^{\rm st}\rangle$ significantly deviate from ${|0\rangle, |1\rangle}$. (iii) For the PCL one, the steady-state populations $P_\pm^{\rm st}$ represented in its diagonalized basis exhibit a large imbalance, yielding a system state much lower entropy than that of the CL model. These features together highlight the non-Markovian and coherence-preserving nature of the PCL environment in contrast to the dissipative CL bath.

Figure \ref{fig2} illustrates the time evolution of the reduced density operator, $\rho_{\tS}(t)$, for both the PCL and CL models. To visualize the system dynamics, we employ the Bloch sphere representation, where each component is defined as $\bar\sigma_i(t) \equiv {\rm tr}{\tS}[\hat\sigma_i\rho{\tS}(t)]$ ($i = x, y, z$). The trajectory of a mixed state is then represented as a path within the unit sphere. Panels (c) and (d) of \Fig{fig2} display the Bloch trajectories for the PCL and CL models, respectively, while panels (a) and (b) present the corresponding transient behaviors of $\bar\sigma_z(t)$ and the Bloch vector magnitude, $|\bar{\bm\sigma}(t)| \equiv \sqrt{\bar\sigma_x^2(t) + \bar\sigma_y^2(t) + \bar\sigma_z^2(t)}$. The two models exhibit qualitatively distinct dynamical features.
(i) At short times, the PCL dynamics show pronounced high-frequency coherent oscillations, whereas the CL dynamics exhibit rapid and nearly monotonic decoherence.
(ii) In the long-time limit, the CL model relaxes to a steady state that is diagonal in the eigenbasis of the system Hamiltonian, while the PCL steady state does not commute with $H_{\tS}$; its eigenvectors, $|\psi_{\pm}^{\rm st}\rangle$, deviate significantly from ${|0\rangle, |1\rangle}$.
(iii) In the PCL case, the steady-state eigenvalues $P_\pm^{\rm st}$ of $\rho_{\tS}^{\rm st}$ display a pronounced imbalance, leading to a much lower entropy compared with the CL model. Together, these observations highlight the non-Markovian and coherence-revival nature of the PCL environment, in sharp contrast to the dissipative character of the CL bath.

\begin{figure}[t]
    % \centering
    \includegraphics[width=0.85\columnwidth]{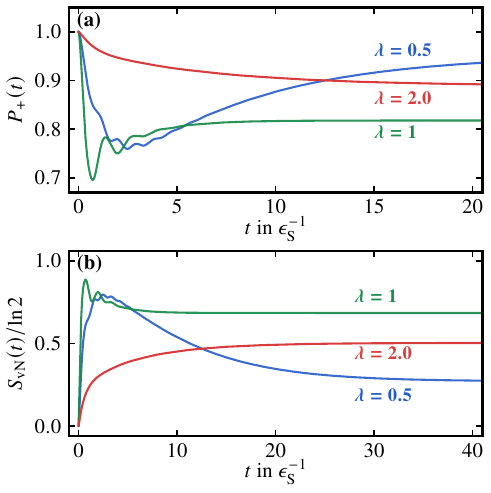}
    \caption{Population of system density operator in the instantaneous eigenbasis and von Neumann entropy calculated with $\lambda = 0.5, 1$, and $2$. Other parameters are given by $\alpha = 2\epsilon_{\tS}$, $\gamma = \epsilon_{\tS}$, $k_BT = 2\epsilon_{\tS}$, and $\xi = 1$. The steady state under the PCL model shows a nonmonotonic dependence on $\lambda$, remaining low entropy in both the weak and strong coupling limits.}
    \label{fig3}
\end{figure}

The above behaviors enlighten us that the exponential system–bath coupling in the PCL model induces a substantial renormalization of the system Hamiltonian. To quantify this effect, we introduce an effective system Hamiltonian $H_{\tS}^{\rm eff}$ (also named as the Hamiltonian of mean force) through system's steady-state \cite{Hil11031110,Ile14032114,Sei16020601},
\begin{align}
\rho_{\tS}^{\rm st} \equiv \lim_{t\to+\infty}\rho_{\tS}(t) \equiv \frac{1}{Z_{\rm eff}} e^{-\beta H_{\tS}^{\rm eff}},
\end{align}
where $Z_{\rm eff} = {\rm tr}{\tS}(e^{-\beta H{\tS}^{\rm eff}})$. 
From the zeroth tier of the equations of motion [\Eq{eom}], a first-order estimation yields
\begin{align}\label{Heff}
H_{\tS}^{\rm eff} \approx H_{\tS} + \langle H_{\SB}\rangle_{\B} = H_{\tS} + 2g\hat S,
\end{align}
with $g = e^{-\lambda^2\langle \hat F^2\rangle_{\B}/2}$ defined in \Eq{eom}.
The energy splitting between the eigenvalues of $H_{\tS}^{\rm eff}$ approximately determines the short-time oscillation frequency of $\rho_{\tS}(t)$,  given by $2\sqrt{\epsilon_{\tS}^2 + 4\alpha^2g^2}$ for the two-level model in \Fig{fig2}. 

\begin{figure}[t]
    % \centering
    \includegraphics[width=0.85\columnwidth]{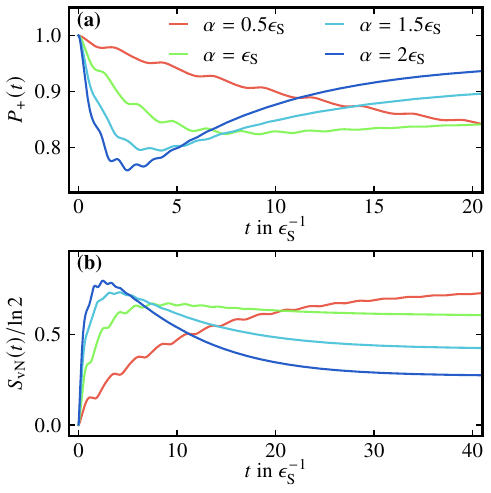}
    \caption{Population of system density operator in the instantaneous eigenbasis and von Neumann entropy calculated with $\alpha = 0.5, 1, 1.5$, and $2\epsilon_{\tS}$. Other parameters are given by $\lambda = 0.5$, $\gamma = \epsilon_{\tS}$, $k_BT = 2\epsilon_{\tS}$, and $\xi = 1$. The equilibrium entropy decreases monotonically when increasing $\alpha$. }
    \label{fig4}
\end{figure}

For a clearer illustration of the influence of the PCL bath, we express the reduced density operator in the instantaneous eigenbasis, $\rho_{\tS}(t) = P_+(t)|\psi_+(t)\rangle\langle\psi_+(t)| + P_-(t)|\psi_-(t)\rangle\langle\psi_-(t)|$, and evaluate the von Neumann entropy, $S_{\rm vN}(t) \equiv -{\rm tr}{\tS}[\rho{\tS}(t)\ln\rho_{\tS}(t)] = -P_+(t)\ln P_+(t) - P_-(t)\ln P_-(t)$, for various values of $\lambda$ (\Fig{fig3}) and $\alpha$ (\Fig{fig4}). As illustrated in \Fig{fig3}, when $\lambda$ is small, the population dynamics display damped oscillations whose frequency decreases with decreasing $\lambda$. In contrast, for sufficiently large $\lambda$, the relaxation becomes monotonic without visible oscillations. The steady-state von Neumann entropy exhibits a nonmonotonic dependence on $\lambda$, remaining low in both the weak and strong coupling limits. Regarding the dependence on $\alpha$, the dynamics consistently show damped oscillations, while both the oscillation frequency and the steady-state entropy vary monotonically with $\alpha$.

\paragraph*{Summary.}
In summary, we have established an exact and nonperturbative framework for the Phase-Coupled Caldeira-Leggett model, unveiling a new class of quantum dissipative dynamics beyond linear system–bath coupling. By employing the dissipaton formalism and introducing a generalized normal ordering for dissipaton operators, we derived a closed, hierarchical set of equations of motion that captures full non-Markovian effects in exponentially coupled environments. The resulting dynamics exhibit rich and distinctive behavior, illustrating how phase-mediated interactions qualitatively alter decoherence and relaxation processes. This framework opens new avenues for exploring quantum transport, strong light–matter coupling, and superconducting or molecular junctions, where phase-type environment couplings are expected to play a decisive role. The methodology developed in this work  provide insights for open other quantum system approaches that are not based on the influence functional formalism, such as generalized quantum master equation \cite{Shi038173, Shi0410647}, stochastic methods \cite{Sha045053}, pseudomode approach  \cite{Gar974636,Tam18030402, Tam19090402,Lam193721,Cir24033083}, and memory kernel coupling theory \cite{Liu25148001,Bi25224106}. 

\vspace{1em}

\paragraph*{Acknowledgments.}
Support from the National Natural Science Foundation of China (Grant Nos.\  224B2305, 22373091, 22173088) and the Innovation Program for Quantum Science and Technology (Grant No.\ 2021ZD0303301) is gratefully acknowledged.

%\bibliography{bibrefs.bib}


\end{document}